# Robustness of Complex Networks Considering Load and Cascading Failure under Edge-removal Attack

Peng Geng[1], Huizhen Hao[1], Annan Yang[2], Yan Liu[3]

1 School of Information and Communication Engineering, Nanjing Institute of Technology, Nanjing 211167, China
2 School of Communication Engineering, Tongda College of Nanjing University of Posts & Telecommunications, Yangzhou 225127, China
3 School of Mathematics and Physics, Nanjing Institute of Technology, Nanjing 211167, China
Correspondence should be addressed to Peng Geng; gengpeng@njit.edu.cn

**Abstract**
In the understanding of important edges in complex networks, the edges with larger degree are naturally considered more important, and they will cause greater destructiveness when attacked. However, through simulation analysis, we conclude that this understanding needs to be based on certain preconditions. In this article, the robustness of BA scale-free network and WS small-world network is studied based on the edge-removal attack strategy considering the edge load and cascading failure. Specific attack methods include: High Load Edge-removal Attacks (HLEA) and Low Load Edge-removal Attacks (LLEA). The simulation results show that the importance of edges is closely related to the load parameter δ. When $0<\delta<1$, attacking the edge with smaller degree will cause greater cascading failure. For this condition, the edge with smaller degree is more important. When $\delta=1$, cascading failure is basically independent of the edge degree. When $\delta>1$, attacking the edge with larger degree will cause greater cascading failure. Therefore, the edge with larger degree is more important for this condition.

## 1. Introduction

At present, with the increasing degree of modernization, various forms of complex networks are ubiquitous in our lives. For example, the complex relationship between individuals constitutes a huge social network. Various forms of transportation have produced multi-layer dependent networks. Complex communication power network infrastructure meets our daily needs. These complex networks play a key role in modern society [1-3]. Therefore, in recent years, there have been more and more analysis and research on real networks based on complex network theory. Amani A M et al. [4] describe the power grid based on the complex network theory and analyze the flexibility and reliability of the network. The benchmark power grid and the actual power grid are simulated. The results show that a small part of the faults in the power grid components may rapidly spread to the whole network, leading to cascading failures, and may even lead to serious power failure of the whole power grid. Cimini G et al. [5] reconstruct the local network based on statistical physical methods and complex network models. An analytical framework to reproduce the local network is proposed. This framework can reconstruct the network structure in the case of incomplete information caused by the failure of a part of the network. Zhang M et al. [6] analyze the congestion of complex traffic networks based on the characteristics of traffic flow. The robustness of partial failure of traffic network caused by congestion is discussed, which provides a theoretical basis for the application of complex network theory in traffic network. It can be seen that when studying real complex networks, the analysis based on partial network faults is the common ground of many articles. Cascading failures often occur in some parts of the network and quickly spread to the whole network, which have great damage to the robustness of the network. In the research of robustness of complex networks, Wang et al. [7] study the robustness of the network considering the cost of attack under different complex network models and different robust performance evaluation indicators, and analyze the robust performance change process. Shi et al. [8] propose a tunable supply chain network for the analysis of mixed cascade faults in undirected supply chain networks, and analyze the robust performance of tunable parameters against random and deliberate attacks in the network model. Yang et al. [9] study the controllability robustness of the network from the perspective of network deliberate attack, propose a layered attack framework, and design corresponding experiments, and put forward suggestions conducive to the robustness of network controllability in the structure of the experiment.

Considering the load-capacity model and cascading failure, this article mainly studies the robustness of complex networks based on edge attacks. Several major aspects are studied in this article, as described below:

- Identify the key elements and relationship of network cascading failure, and build a general analysis framework of network cascading failure.
- The initial load is defined based on the characteristics of edges and the load redistribution strategy considering capacity is considered.
- The dynamic network cascading failure model is studied, and the actual load distribution process of BA scale-free network and WS small-world network with cascading failure is simulated.
- The invulnerability of different networks against cascading failures under different parameters is compared, which is helpful to select the appropriate network topology and the value of the adjustable parameters that match it in the future.

The rest of this article is organized as follows. The current state of research on cascading failure of complex networks is presented in Section 2. Section 3 gives the network model and describes the construction process of BA scale-free network and WS small-world network. Section 4 analyzes the network cascading failure model, in which the parameter control and failure process are described. Two attack strategies (HLEA and LLEA) are proposed in Section 5, and robustness metrics are given. Section 6 carries out simulation evaluation. The summary of the article comes from section 7. Section 8 describes the future work.

## 2. Related Work

In the past, factors such as load, capacity and load redistribution have been used to build cascading failure models of networks. The research objects include nodes and edges. This section discusses some of the past works that has been completed recently.

### 2.1. Research on Node Model

Node-based cascading failure model is the most basic research. In [10], nodes in the network are divided into different states, and the state transition process is given according to the failure type. In addition, when the load of a node exceeds its capacity, it will not fail immediately, but wait for a time slice before deleting the node. This delay mode is applicable to complex communication networks, but it is questionable whether it is applicable to the power grid with node failure immediately after overload. Literature [11] studies the cascading failure characteristics of coupling networks on the basis of considering node degree and average node degree. Different from the previous homogeneous coupling networks, this paper constructs a multi-subnet composite complex network through different types of networks. The research shows that the greater the coupling between the layers of networks, the stronger the robustness. However, the dependency between subnets in this paper is unidirectional, which is inconsistent with the bidirectional dependency of real networks. Literature [12] studies the vulnerability of smart grid, and points out that attacking nodes with high load in the network does not necessarily achieve a large degree of damage. This article puts forward MICLLB attack strategy from the perspective of attackers, and uses HEBDA defense strategy to conduct attack and defense drills. In addition, the article also proves that this kind of attack is NP-complete problem, and the defense strategy performs well in the range of affordable computational complexity. A cascading failure model based on Sink node is proposed in literature [13]. It designs the network topology optimization algorithm MA-TOSCA to increase the network robustness in a shorter time. In addition, the simulation also proves that the onion network has better robustness. Literature [14] regards enterprises as nodes and studies the cascading failures of supply chain networks. Enterprise nodes have upper and lower capacity. It is proved that the upper limit capacity is negatively correlated with cascading failure, and the lower limit capacity is positively correlated with cascading failure. In fact, cascading failure mainly depends on the lower limit capacity, and the determination of its specific value helps to prevent cascading failure in the supply chain network.

**2.2. Research on Edge Model**

Literature [15] proposes a method to predict the critical edge of power grid. The research shows that the flow of power grid can cause cascading failure in the second level transient. The length between the current failure edge and the failure source is also used to measure the magnitude of the failure details. In this way, the propagation distance and propagation time of network cascading failures can be observed, and further, a real-time scheme to stop or control cascading failures is designed. A research model of edge cascading faults in power systems was proposed by literature [16]. This model makes statistics on the historical cascading failure data, and makes machine learning prediction based on Bayesian network to obtain the edge most likely to have cascading failure. Research shows that machine learning simplifies the calculation of complex networks. Literature [17] assumes that attackers can only detect some edges in complex networks. Random attacks and target attacks are implemented in the article. In the network model, the scale-free network is adopted and the edge is weighted by the medium. The simulation results show that partial edge attacks will reduce network connectivity. In addition, the article also gives the network protection strategy. The robustness of the community network edge attack strategy is analyzed by literature [18]. This article proposes the MA-Rcom algorithm to enhance the edge robustness of the community network. Simulation results in real networks show that MA-Rcom algorithm has potential value in solving cascading failures.

**2.3. Research on Hybrid Model**

The cascading failure characteristics of scale-free networks based on node benefits are studied in literature [19]. In this paper, nodes are divided into normal nodes, negative benefits nodes and failure nodes. In the process of network operation, the edge between the negative benefits node and the failure node is deleted, and the cascading failure process of the complex network is analyzed. Literature [20] considers both node capacity and edge capacity, and studies the cascading failure characteristics of complex wireless sensor networks. In this paper, a route recovery mechanism is proposed by setting cache for nodes in the network and bandwidth for edges. Nodes and edges that fail can be recovered by this mechanism. The simulation results show that the network survivability has a strong correlation with node cache. Based on multi-sink wireless sensor networks, a hybrid cascading failure model based on nodes and edges is proposed in literature [21]. Experiments show that there is a critical threshold between capacity (node capacity and edge capacity) and the load (node load and edge load) in the network. Increasing capacity to break through the threshold impact can effectively curb the damage of cascading failures. Literature [22] abstracts the urban traffic network as a two-level coupling network of public transit and subway, and gives a certain capacity to the nodes and edges in the network. Based on the measurement of topology and flow, the vulnerability calculation method of the traffic network is given. The research shows that the cascading failure of urban traffic network does not necessarily increase with the increase of bad weather, and there should be the intervention factor of traffic management department.

## 3. Network Models

Graph is a unified tool to describe the network, so as to study the topology of different complex networks. Specifically, the network can be abstracted as a graph composed of a set of points V and a set of edges E ( $G=(V,E)$ ), where $V=\{v_i\,|\,i=1,2,...,n\}$ represents a set of n nodes in the network, $E=\{e_k\,|\,e_k=(v_i,v_j),k=1,2,...,p\}$ represents a set of connections between nodes in the network, $e_{ij}\neq 0$ represents at least one edge between node i and node j. The number of nodes in the network represents the size of the network. Each edge in E corresponds to a pair of nodes in V. If the edges $e_{ij}$ and $e_{ji}$ between any two nodes i and node j are the same edge, the network is called an undirected network, otherwise it is called a directed network. If each edge is given a corresponding weight value, the network is called a weighted network, otherwise it is called an unweighted network. The networks studied in this paper can be abstracted as undirected weighted networks.

Some typical network topology models are used to replace the real system, so as to analyze the process and consequences of network cascading failure, and better understand the network disaster. At present, typical network topology models include regular network, random network, small-world network and scale-free network. Among them, regular networks cannot completely describe large-scale networks in reality, so complex networks are no longer used in the research. The random network has unknown organization rules, so the ER random network model is rarely used in the research of real networks. Many networks of real systems have large clustering coefficients and small average path length, which is the so-called small-world effect. Therefore, small-world network models are also commonly used in the study of complex networks. The WS small-world model is the most widely used in small-world network models [23]. Because many networks in real life have scale-free characteristics, BA scale-free network is also the most commonly used network model in the study of complex networks [24]. Based on the above analysis, this paper selects WS small-world network and BA scale-free network as the network models to discuss cascading failures.

The construction algorithm of WS small-world model is as follows:

1) Start from the rule graph: considering a nearest neighbor coupling network with N nodes, which is surrounded by a ring, where each node is connected to its left and right adjacent K/2 nodes, and K is an even number.

2) Randomized reconnection: randomly reconnects each edge in the network with probability p, that is, one endpoint of the edge remains unchanged, and the other endpoint is taken as a randomly selected node in the network. It is stipulated that there can be at most one edge between any two different nodes, and each node cannot have an edge connected to itself.

The construction algorithm of BA scale-free model is as follows:

1) Initialization and growth: Start from a network with $m_0$ nodes, introduce a new node each time, and connect to $m$ existing nodes until the number of nodes reaches the predetermined N.

2) Priority connection: The probability $\prod_i$ of a new node connecting with an existing node i, the degree $k_i$, and the degree $k_j$ satisfy the relationship $\prod_i = \dfrac{k_i}{\sum_{j=1}^{N} k_j}$.

## 4. Cascading Failure Model

The attack on the network is generally caused by external interference, while the failure node (edge) removal strategy, the failure node (edge) load redistribution strategy, the load distribution and capacity distribution caused by different routing rules belong to the internal mechanism of the network. Both the external interference of the network and the adjustment of the internal mechanism will cause different degrees of cascading failures, which will ultimately affect the connectivity and coverage of the network. How much cascading failure will occur due to the fracture of the edge caused by external interference and internal mechanism? This is the primary issue of this paper. The parameters involved in describing cascading failures are explained in Table 1.

*Table 1. Parameter Interpretation*

| Symbol | Meaning |
|---|---|
| $L_{ij}$ | The initial load of the edge $e_{ij}$ |
| $C_{ij}$ | The edge capacity of the edge $e_{ij}$ |
| $\delta$ | Load parameter |
| $k_i$ | Degree of node $i$ |
| $\varepsilon$ | Capacity parameter |
| $\theta$ | Capacity parameter |
| $P_{ia}$ | The proportion of additional load |
| $\tau_i$ | The set of neighbor nodes of $i$ |
| $\Delta L_{ia}$ | The increased load on edge $e_{ia}$ |
| $M$ | The number of edges |
| $A$ | The set of attacked edges |
| $M_{e_{ij}}$ | The number of failure edges when edge $e_{ij}$ is attacked |
| $\Gamma$ | The network robustness measurement |

### 4.1. Limitation Factors

The understanding of cascading failure processes will depend on the following definitions.

- Initial Load: The edge between node $i$ and node $j$ is $e_{ij}$. Therefore, we define the initial load $L_{ij}$ of the edge $e_{ij}$ at a certain moment in the network as

$$L_{ij} = (k_i k_j)^\delta. \qquad (1)$$

Where $k_i$ and $k_j$ respectively represent the degrees connecting the two endpoints $i$ and $j$ of edge $e_{ij}$. The parameter $\delta(\delta > 0)$ is load parameter. By adjusting the value of $\delta$, the strength of $L_{ij}$ can be controlled, and different network load distributions in the actual network can be simulated at the same time.

- Edge Capacity: Edge capacity expresses the maximum load that the edge can bear. We define the edge capacity $C_{ij}$ of the edge $e_{ij}$ in the network as

$$C_{ij} = L_{ij} + \varepsilon L_{ij}^\theta. \qquad (2)$$

Where the parameter $\varepsilon(\varepsilon > 0)$ and $\theta(\theta > 0)$ are capacity parameters. By adjusting the capacity parameters, the relationship between network capacity and load in different proportions can be obtained. Obviously, the relationship between $C_{ij}$ and $L_{ij}$ is linear when $\theta = 1$. The subsequent simulations in this article mainly show the relationship between $\delta$ and $\varepsilon$. For convenience, we consider the case of $\theta = 1$, as shown in the following equation.

$$C_{ij} = (1+\varepsilon)L_{ij} \qquad (3)$$

- Proportion of Load Redistribution: When $e_{ij}$ is attacked or fails due to the excessive load on it, the original load on it will be redistributed to the neighbor edges whose endpoints contain node $i$ or $j$. There are two ways of load redistribution. One is equal distribution, that is, neighbor edges evenly bear the load released by $e_{ij}$. The other is non-uniform allocation, that is, allocation is based on a certain proportional preference. This article adopts the latter scheme. For a neighbor edge $e_{ia}$ of $e_{ij}$, the proportion of additional load to be loaded is $P_{ia}$, which is defined as shown in the following equation.

$$P_{ia} = \frac{C_{ia} - L_{ia}}{\sum_{b \in \tau_i, j \notin \tau_i}(C_{ib} - L_{ib}) + \sum_{d \in \tau_j, i \notin \tau_j}(C_{jd} - L_{jd})}$$
$$= \frac{(k_i k_a)^\delta}{\sum_{b \in \tau_i, j \notin \tau_i}(k_i k_b)^\delta + \sum_{d \in \tau_j, i \notin \tau_j}(k_j k_d)^\delta} \qquad (4)$$

Where $\tau_i$ represents the set of neighbor nodes of $i$ (excluding $j$), and $\tau_j$ represents the set of neighbor nodes of $j$ (excluding node $i$).

- Increased Load: After the failure of connection edge $e_{ij}$, the increased load (denoted as $\Delta L_{ia}$) of its neighbor edge $e_{ia}$ is proportional to initial load, as shown in the following equation.

$$\Delta L_{ia} = L_{ij} \times P_{ia}$$
$$= (k_i k_j)^\delta \times \frac{(k_i k_a)^\delta}{\sum_{b \in \tau_i, j \notin \tau_i}(k_i k_b)^\delta + \sum_{d \in \tau_j, i \notin \tau_j}(k_j k_d)^\delta} \qquad (5)$$

It can be seen that in the process of load redistribution, the neighbor edge with larger capacity will get more load, which can effectively reduce further possible cascading failures. Figure 1 briefly describes the process of load reloading. When $e_{ij}$ fails (shown by the dotted line in the Figure 1), the load $L_{ij}$ on it is distributed to neighbors $e_{ia}$, $e_{ib}$, $e_{ic}$ and $e_{jd}$. Thus, the loads on $e_{ia}$, $e_{ib}$, $e_{ic}$ and $e_{jd}$ are increased from $L_{ia}$, $L_{ib}$, $L_{ic}$ and $L_{jd}$ to $L_{ia} + \Delta L_{ia}$, $L_{ib} + \Delta L_{ib}$, $L_{ic} + \Delta L_{ic}$ and $L_{jd} + \Delta L_{jd}$.

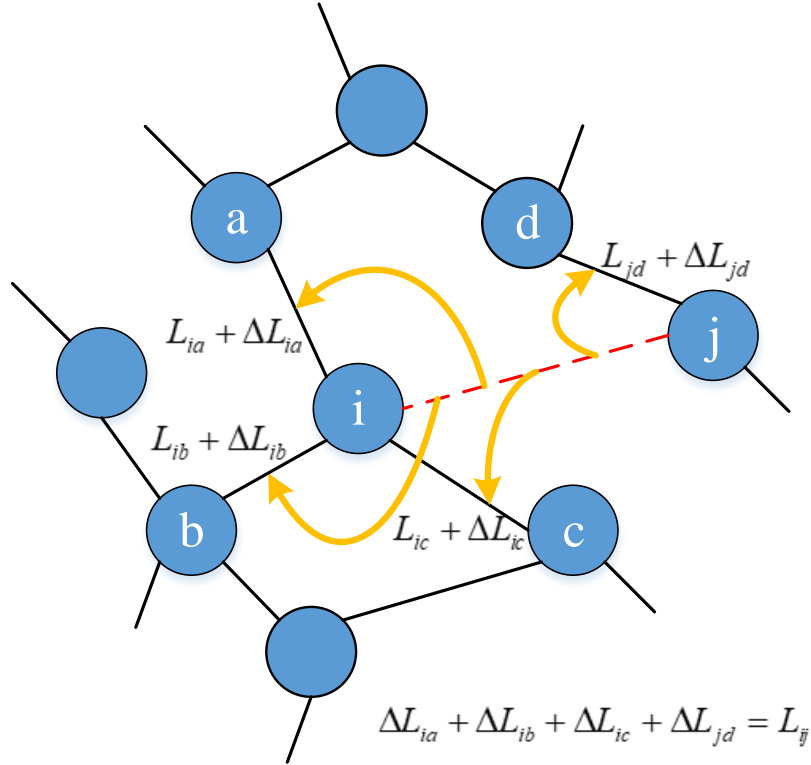

*Figure 1. Load redistribution*

### 4.2. Cascading Failure Process

In real networks, especially in communication networks, when a communication line fails, the solution is to use backup equipment to reconnect the failed link. However, due to the large scale of the network, in view of the cost problem, only some important links can support the backup method in the network topology. Therefore, we focus on establishing cascading failure propagation models for most networks without backup equipment.

Figure 2 describes the one step cascading failure process. The left side of the Figure 2 expresses the load redistribution process after the communication link $e_{ij}$ is attacked and fails. It can be seen that the load on the communication link $e_{jd}$ has increased to $L_{jd} + \Delta L_{jd}$. If the load at this time exceeds its capacity ($L_{jd} + \Delta L_{jd} > C_{jd}$), $e_{jd}$ will also fail. Thus, another load redistribution behavior occurs, as shown on the right side of Figure 2. This creates the one step cascading failure (due to the failure of $e_{ij}$, which leads to the failure of $e_{jd}$). If this effect is passed on, the cascading failure of the whole network may occur, resulting in network paralysis.

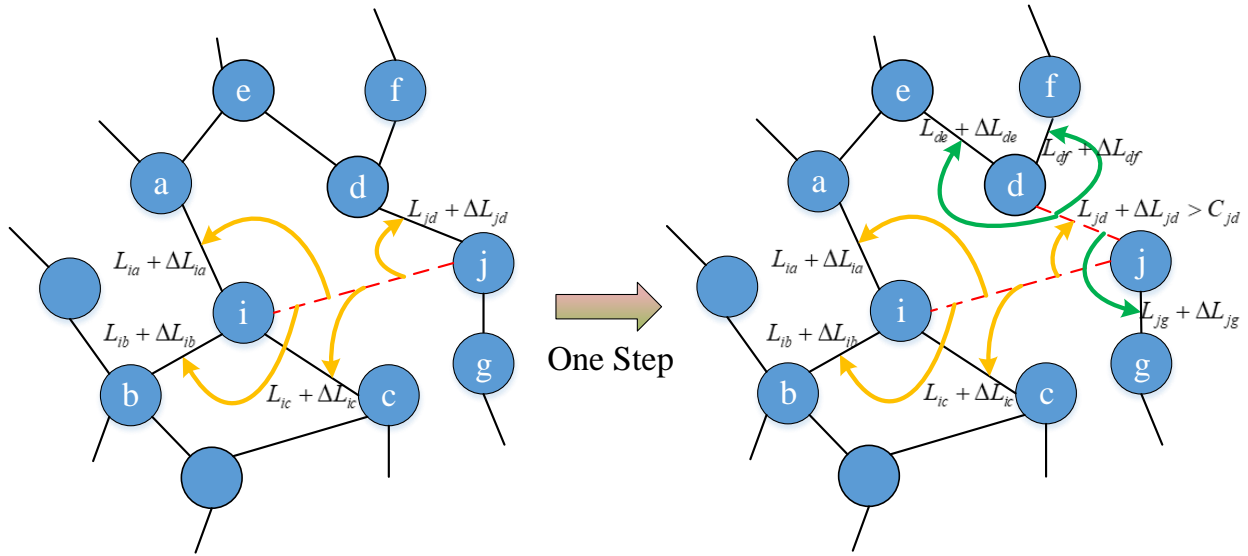

*Figure 2. One step cascading failure process*

In order to facilitate the subsequent simulation process, we can describe the network operation and parameter calculation process as the following steps. The flow chart expression is shown in Figure 3.

1) Generate a fully connected undirected complex network topology with a certain network model.
2) Calculate the degree of each node, calculate the load of each edge based on this, and store it in the set $L$. Determine the parameter $\varepsilon$, calculate the capacity of each edge, and store it in the set $C$.
3) Attack the network based on the edge-removal attack strategy.
4) Scan the network to find isolated nodes and remove them.
5) Start the load redistribution process on the removed edge. Calculate the added load $\Delta L$ on the neighbor edges.
6) Judge whether there is a situation of $L + \Delta L > C$, if so, return to step 3). If it does not exist, go to step 7).
7) The network runs a life cycle to the end.

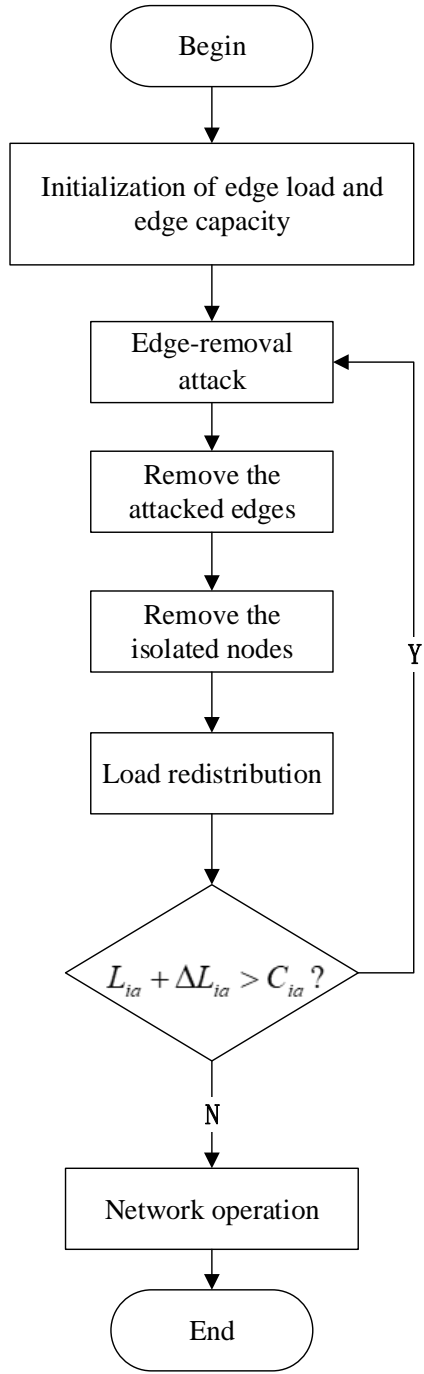

*Figure 3. The flow chart of network operation*

## 5. Edge-removal Attack

In this article, two types of edge-removal attacks will be carried out for different network models, including High Load Edge-removal Attacks (HLEA: removing the edge with a higher load in the network) and Low load Edge-removal Attacks (LLEA: removing the edge with a lower load in the network). Firstly, we need to sort the load of the edges after initializing the network.

Secondly, the edges with the higher (lower) load are selected for removal in each round of attack. Finally, if there are multiple edges with the same load, we will randomly select one of the periods.

The number of edges in the network is denoted as $M$, the set of attacked edges is denoted as $A$, $M_A$ represents the number of attacked edges, and $M_{e_{ij}}$ represents the number of cascading failure edges generated after edge $e_{ij}$ is attacked. Therefore, a network robustness index $\Gamma$ to measure the cascading failure is proposed, as shown in the following equation.

$$\Gamma = \frac{\sum_{e_{ij} \in A} M_{e_{ij}}}{M_A(M-1)} \qquad (6)$$

Obviously, $\Gamma$ represents the proportion of the number of cascading failure edges caused by each attacked edge in the set $A$ to the total number of edges. This article studies the relationship between $\varepsilon$ and $\Gamma$ for complex networks under HLEA and LLEA attacks when $\delta$ takes a certain range of values.

## 6. Simulation Evaluation

The research idea based on the simulation method is to observe the change of network function when the connection edge fails, and measure the cascading failure degree of the network by the critical threshold when the network state changes. In this paper, when the load parameter $\delta$ takes a certain range, the relationship between the capacity parameter $\varepsilon$ and the network robustness measurement $\Gamma$ is observed. The network topology models are based on WS-small world and BA scale-free. Network attack strategies include HLEA and LLEA. The parameter types and values involved in the simulation are shown in Table 2. The results of each experiment are the average of 30 experiments.

*Table 2. Parameter value setting*

| Type | Value |
|---|---|
| $N$ | 1000 |
| $M_A$ | 10 |
| Parameter $K$ in WS small-world network | 4 |
| Parameter $p$ in WS small-world network | 0.1 |
| Parameter $m_0$ in BA scale-free network | 2 |
| Parameter $m$ in BA scale-free network | 2 |
| $\delta$ | 0.2, 0.4, 0.6, 0.8, 1.0, 1.2, 1.4, 1.6, 1.8, 2.0 |
| $\theta$ | 1.0 |

### 6.1. Attack destructiveness when $0 < \delta < 1$

Since $C_{ij} = (1+\varepsilon)L_{ij}$, it can be seen that the parameter $\varepsilon$ determines the capacity of the edges in the network. When $\varepsilon$ is large enough, the failure of one edge will not cause the failure of its neighbors, that is, there will be no cascading failure. As can be seen from Figure 4 (a) - Figure 4 (d), when $\varepsilon$ decreases to a certain value, the value of $\Gamma$ will rise rapidly, that is, the proportion of cascading failure edges in the network will increase rapidly. Therefore, we only need to observe the value of $\varepsilon$ when $\Gamma$ is 0 to determine which attack strategy is more destructive. The larger the value of $\varepsilon$, the greater destructive. Figure 4 (a) - Figure 4 (d) shows the cascading failure results of different networks under LLEA attack and HLEA attack when $0<\delta<1$. For the aspect, we take the value of $\delta$ as 0.2-0.8. It can be seen that the LLEA attack strategy will lead to greater cascading failure for both BA scale-free networks and WS small-world networks (although with the increase of $\delta$, the $\varepsilon$ corresponding to $\Gamma=0$ becomes closer and closer). That is, attacking the edge with a small degree value brings greater destructiveness. This overturns people's understanding of the important edges in complex networks. People usually think that the edges with larger degree values are more important. However, this experiment shows that when $0<\delta<1$, the edges with smaller degree values are more important and need to be protected.

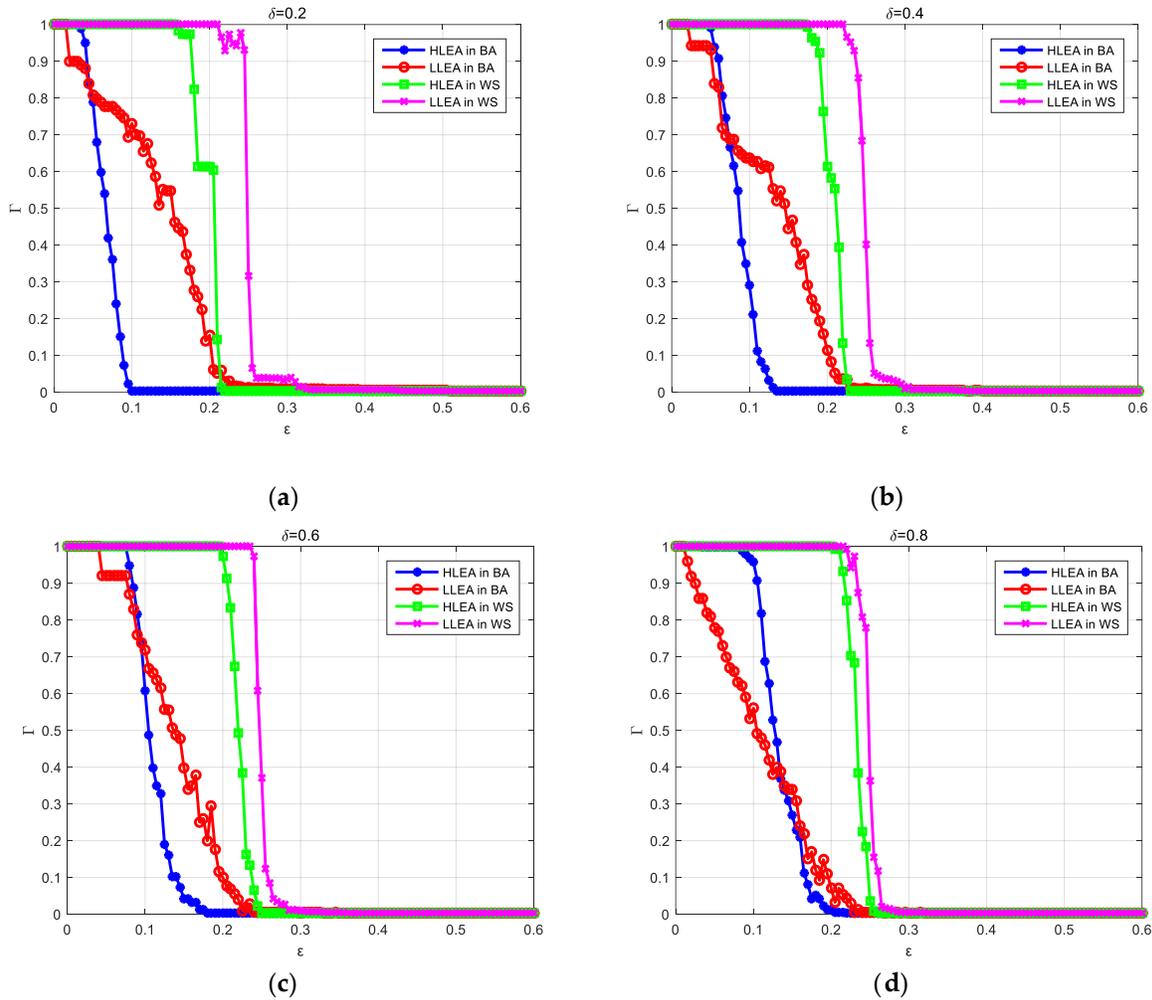

*Figure 4. Cascading failure under LLEA and HLEA attacks when $0<\delta<1$*

**6.2. Attack destructiveness when $\delta \geq 1$**

Figure 5 (a) - Figure 5 (f) are the simulation results of different cascading failures under LLEA attack and HLEA attack when $\delta \geq 1$. In order to analyze the change trend, we take the value of $\delta$ between 1.0 and 2.0. It can be seen that the attack graph based on HLEA gradually moves to the right, that is, the $\varepsilon$ under the same $\Gamma$ is getting larger and larger. On the contrary, in LLEA, the $\varepsilon$ under the same $\Gamma$ is getting smaller and smaller. This shows that when $\delta \geq 1$, whether it is a BA scale-free network or a WS small-world network, attacking the edge with a higher degree will bring greater destruction. This result is consistent with people's common sense, that is, the edge with larger degree value plays a more important role in the network. Of course, the simulation results in this paper show that this requires a prerequisite, that is, $\delta \geq 1$.

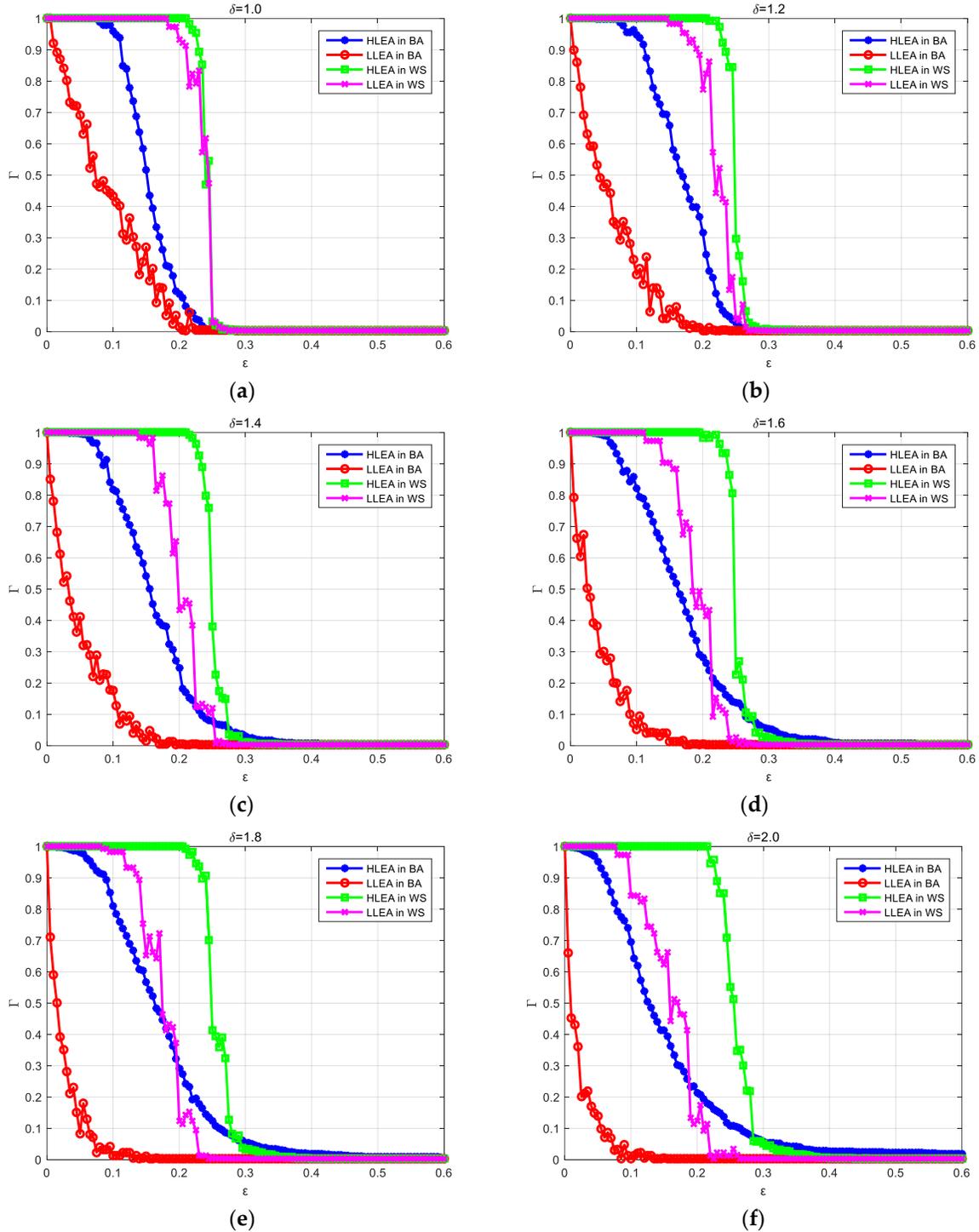

*Figure 5. Cascading failure under LLEA and HLEA attacks when $\delta \geq 1$*

## 6.3. Threshold-based analysis

We record the value of $\varepsilon$ when $\Gamma$ is just 0 as the threshold $\varepsilon_T$, which represents the value of $\varepsilon$ corresponding to cascading failure when a network is attacked. For a network with controllable edge capacity, the larger the capacity of set edge, the higher the cost. Therefore, the edge capacity corresponding to $\varepsilon_T$ is the best capacity. By observing $\varepsilon_T$, we can intuitively observe the impact of network attacks.

In the case of attacking two networks with HLEA and LLEA, the values of $\varepsilon_T$ corresponding to different $\delta$ are shown in Table 3. The change of $\varepsilon_T$ is shown in Figure 6. It can be seen that under the same network, when $0<\delta<1$, the $\varepsilon_T$ corresponding to LLEA attack strategy is always greater than HLEA. That is, attacking edges with smaller degrees will have greater impact. When $\delta=1$, the $\varepsilon_T$ corresponding to LLEA-based attacks and HLEA-based attacks is very close. That is, the attack effect has nothing to do with the edge degree. When $\delta>1$, the $\varepsilon_T$ corresponding to HLEA attack strategy is gradually larger than LLEA. Under this condition, the impact of attacking edges with larger degree is greater.

*Table 3. The value of threshold $\varepsilon_T$*

| $\delta$ | $\varepsilon_T$ 0f HLEA in BA | $\varepsilon_T$ 0f LLEA in BA | $\varepsilon_T$ 0f HLEA in WS | $\varepsilon_T$ 0f LLEA in WS |
|---|---|---|---|---|
| 0.2 | 0.1 | 0.245 | 0.22 | 0.325 |
| 0.4 | 0.135 | 0.235 | 0.23 | 0.3 |
| 0.6 | 0.18 | 0.245 | 0.25 | 0.285 |
| 0.8 | 0.205 | 0.23 | 0.255 | 0.285 |
| 1.0 | 0.24 | 0.225 | 0.265 | 0.275 |
| 1.2 | 0.26 | 0.215 | 0.285 | 0.27 |
| 1.4 | 0.36 | 0.195 | 0.315 | 0.265 |
| 1.6 | 0.4 | 0.175 | 0.34 | 0.255 |
| 1.8 | 0.48 | 0.14 | 0.36 | 0.245 |
| 2.0 | 0.6 | 0.1 | 0.38 | 0.265 |

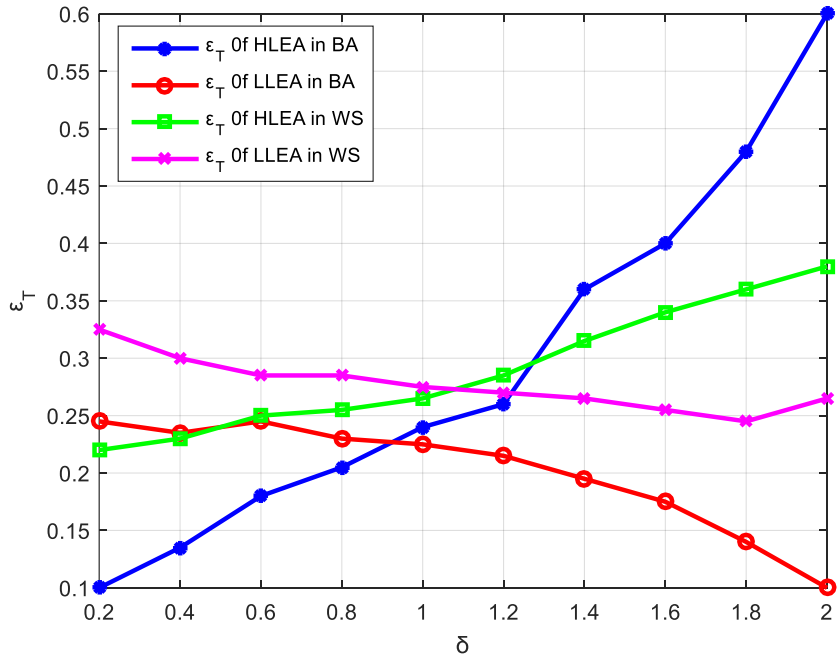

*Figure 6. The changes of $\varepsilon_T$ corresponding to different $\delta$*

## 7. Conclusion

Based on the analysis and summary of related research, this article studies the robustness of complex networks. Firstly, the object of study is edges rather than nodes. Secondly, we consider the load and capacity of the network connection edges, which is more in line with the characteristics of the actual network. Thirdly, the network attack strategy adopts the edge removal method. Fourthly, in terms of measuring the robustness of the network, we adopt the index of the ratio of the number of edges with cascading failures to the total number of edges. Based on the WS small-world network and the BA scale-free network, we analyze the model that the connection edge redistributes the load due to attack and causes cascading failure. Based on two attack strategies, High Load Edge-removal Attacks (HLEA: removing the edge with a high load in the network) and Low Load Edge-removal Attacks (LLEA: removing the edge with a lower load in the network), we conducted simulation analysis in MATLAB. The results show that when the edge load parameter $0<\delta<1$, the LLEA attack strategy is more effective, which overturns people's conventional thinking. When $\delta=1$, there is a linear relationship between load and degree. At this time, the attack strategies of HLEA and LLEA are independent of the cascading failure model. When $\delta>1$, HLEA attack strategy is more effective, which is in line with people's conventional thinking. In addition, this article compares the invulnerability of different networks against cascading failures under different parameters, which is helpful to select the appropriate network topology and the value of the adjustable parameters that match it in the future, so as to achieve the best invulnerability of the network against cascading failures.

## 8. Future Work

The empirical research in the real network is the work that this paper needs to do in the future. For example, interdependent power grids and communication networks, intelligent transportation complex networks, and the Internet of Things can be abstracted from their topological structure and operation process. Robustness analysis can be carried out using the cascading failure model in this paper.

**Data Availability**

Some or all data, models, or codes generated or used during the study are available from the corresponding author by request.

**Conflicts of Interest**

The authors declare that they have no conflicts of interest.

**Author Contributions**

Conceptualization, P.G. and A.Y.; methodology, H.H. and Y.L.; software, P.G. and H.H; writing—original draft preparation, P.G. and A.Y.; writing—review and editing, Y.L. and P.G. We confirm that all contributors and organizations to the manuscript have been listed, and there are no other authors.

**Acknowledgments**

This work was supported by the National Natural Science Foundation of China (grant number 41972111) and the Second Tibetan Plateau Scientific Expedition and Research Program (STEP) (grant number 2019QZKK020604).